\documentclass[letterpaper]{article}
\usepackage{url}
\usepackage{aaai}
\usepackage{times}
\usepackage{helvet}
\usepackage{courier}
\usepackage{url}
\begin{document}
%

\title{\#Election2020: The First Public Twitter Dataset \\ on the 2020 US Presidential Election}

\author{Emily Chen, Ashok Deb, Emilio Ferrara\\
University of Southern California, Information Sciences Institute\\
4676 Admiralty Way, \#1001\\ 
Marina del Rey, CA 90292\\
Corresponding author: emiliofe@usc.edu
}

\maketitle
\begin{abstract}
\begin{quote}


The integrity of democratic political discourse is at the core of guaranteeing free and fair elections. With social media often dictating the tones and trends of politics-related discussion, it is paramount to be able to study online chatter, especially in the run up to important voting events, like the upcoming November 3, 2020 U.S. Presidential Election. Limited access to social media data is often the first barrier to impede, hinder, or slow down progress, and ultimately our understanding of online political discourse. To mitigate this issue and try to empower the Computational Social Science research community, we decided to publicly release a massive-scale, longitudinal dataset of U.S. politics- and election-related tweets. This multilingual dataset that we have been collecting for over one year encompasses hundreds of millions of tweets and tracks all salient U.S. politics trends, actors, and events between 2019 and 2020. It predates and spans the whole period of the Republican and Democratic primaries, with real-time tracking of all presidential contenders of both sides of the aisle. The dataset then focuses on presidential and vice-presidential candidates. Our dataset release is curated, documented and will be constantly updated on a weekly-basis, through the November 3, 2020 election and beyond.  
We hope that the academic community, computational journalists, and research practitioners alike will all take advantage of our dataset to study relevant scientific and social issues, including problems like misinformation, information manipulation, interference, and distortion of online political discourse that have been prevalent in the context of recent election events in the United States and worldwide.

Our dataset is available at: \textbf{\url{https://github.com/echen102/us-pres-elections-2020}} 


\end{quote}
\end{abstract}

\section{Introduction}
\noindent The United States' constitution stipulates that the president serves a 4 year term, with a maximum of two terms. In 2016, Republican candidate Donald Trump and Democratic candidate Hillary Clinton both ran for the presidency. Donald Trump won, becoming the 45th President of the United States, beating Hillary Clinton in electoral votes 306 to 232; Clinton, however, won the popular vote by almost 3 million more votes.\footnote{\url{https://www.nytimes.com/elections/2016/results/president}} Four years later, it is again time for the United States to return to the polling booths (or, in the current times, to the mail box) to vote for the individual who will be the president of United States for the coming 4 years: incumbent Republican Donald Trump or the Democratic challenger, and former Vice-President Joe Biden. 

Historically, the incumbent president is favored to win the nomination for their party's nominee for the president of the United States;\footnote{\url{https://time.com/5682760/incumbent-presidents-primary-challenges/}} although Trump did face a few challengers from the Republican party, it became increasingly clear that he would gain the Republican party's nomination. 

The Democratic primary, however, was a contentious race, eliciting one of the largest candidate pools in modern American politics, with 29 candidates vying to be the Democratic party's nominee for president.\footnote{\url{https://www.politifact.com/article/2019/may/02/big-democratic-primary-field-what-need/}} The large pool of candidates necessitated the Democratic debate to be held on two separate nights in order to accommodate the candidates, but as time wore on, many candidates began to drop out of the race. The advent of COVID-19 in the United States in March 2020, and the ensuing regulations to encourage social distancing, forced the remaining campaigns, in particular Joe Biden's and Bernie Sander's campaigns, to shift to a virtual campaign model. On April 8, 2020, Bernie Sanders dropped out, leaving Joe Biden as the presumptive Democratic party presidential nominee. 

Joe Biden announced his selection of Kamala Harris as his running mate on August 11, 2020, and he officially accepted the Democratic nomination on August 20, 2020, during the Democratic National Convention. Donald Trump officially accepted his nomination on August 27, 2020, during the Republican National Convention. 

As the final sprint to election day on November 3, 2020 begins, Americans are taking to online social platforms to voice their opinions and engage in conversation surrounding the upcoming elections. Twitter has historically been a platform used by politicians to reach their base \cite{jungherr2016twitter}. 

Inspired by the impact that our similar initiative to share a COVID-19 Twitter dataset had \cite{chen2020tracking}, in this paper, we briefly document the first public release of our election-related dataset that we have been collecting for over one year.  We hope that, in releasing this dataset, the research community can leverage its content to study and understand the dynamics in a highly contentious election held during a pandemic, particularly with reports of confirmed foreign interference already surfacing.\footnote{\url{https://home.treasury.gov/news/press-releases/sm1118}}

\section{Data Collection}

We uninterruptedly collected election-related tweets beginning \textbf{May 20, 2019}, and have continued collection efforts since then. We use Twitter's streaming API through Tweepy
and follow specific mentions and accounts related to candidates who were running to be nominated as their party's nomination for president of the United States, in addition to a manually-compiled, general election-related list of keywords and hashtags. As candidates officially announced the suspension of their campaigns, their respective accounts and mentions were removed from our real-time tracking list. However, for a subset of these accounts, we decided to restart tracking at later dates in the future, for reasons associated to real-world events, most notably political events, in addition to adding supplemental keywords and accounts to our tracking list. This process is documented in Table~\ref{mention_table}.

We will continue to collect election-related tweets through the elections and for a few months after the presidential-elect is declared (depending on when votes are all counted due to the mail-in ballots), so as to capture the nation's activity during the election season and the reaction to the result. We have collected well over 600 million tweets, resulting in over 4 TB of raw data. Our first release is of tweets from 6/20/2020 through 9/06/2020, constituting about 240 million tweets and almost 2 TB of raw data. In future releases, we will continue processing and adding data we collected prior to 6/20/2020 and after 9/06/2020. We anticipate that our entire data collection will contain well over one billion tweets, as the data keeps growing rapidly as we get closer to the November 3, 2020 election.

\textbf{Note}: Twitter's Developer Agreement \& Policy stipulates that we are unable to share any data specific to individual tweets except for a tweet's Tweet ID. As a result, we are releasing a collection of Tweet IDs that researchers are then able to use in tandem with Twitter's API to retrieve the full tweet payload. We recommend using tools such as DocNow's Hydrator\footnote{\url{https://github.com/DocNow/hydrator}} or Twarc;\footnote{\url{https://github.com/DocNow/twarc}} we do note that if tweets have been deleted from Twitter's platform, researchers will be unable to retrieve the payloads for those tweets. In our repository, we provide ready-to-use Python code scripts to perform all the operations described above.

\subsection{Tracked Keywords and Accounts}

In order to capture the discourse surrounding the elections, we followed specific user mentions and accounts that are tied to the official and personal accounts of candidates running for president. Twitter's streaming API gives us access to approximately 1\% stream of all tweets in real-time, and takes in a list of keywords and returns any tweet within that sample stream that contains any of the keywords in the metadata and text of the tweet payload. This results in us not needing to track every permutation of each keyword. We list a sample of the mentions and accounts that we tracked in Table \ref{mention_table} and a sample of the keywords we tracked in Table \ref{keyword_table}. A full list can be found in the accounts.txt file and keywords.txt file in our data repository. 

\begin{table}[t]
    \centering
    \footnotesize
    \begin{tabular}{cccc}
    \textbf{Mentions} & \textbf{Started Tracking} & \textbf{Stopped} & \textbf{Restarted}\\
    \hline
@realDonaldTrump & 5/20/19 & - & - \\ 
@GovBillWeld & 5/20/19 & - & -  \\ 
@MarkSanford & 5/20/19 & 11/14/19 & 9/25/20 \\
@WalshFreedom & 5/20/19 & - & -  \\ 
@MichaelBennet & 5/20/19 & - & -  \\ 
@JoeBiden & 5/20/19 & - & -  \\ 
@CoryBooker & 5/20/19 & 1/13/20 & 9/25/20 \\
@GovernorBullock & 5/20/19 & 12/2/19 & 9/25/20 \\
@PeteButtigieg & 5/20/19 & - & -  \\ 
@JulianCastro & 5/20/19 & 1/2/20 & 9/25/20 \\
@BilldeBlasio & 5/20/19 & 11/14/19 & 9/25/20 \\
@JohnDelaney & 5/20/19 & - & -  \\ 
@TulsiGabbard & 5/20/19 & - & -  \\ 
@gillbrandny  & 5/20/19 & 11/14/19 & 6/20/20 \\
@KamalaHarris & 5/20/19 & 12/3/19 & 6/20/20 \\
@SenKamalaHarris & 5/20/19 & 12/3/19 & 6/20/20 \\
@Hickenlooper & 5/20/19 & 11/14/19 & 9/25/20 \\
@JayInslee & 5/20/19 & 11/14/19 & 9/25/20 \\
@amyklobuchar & 5/20/19 & - & -  \\ 
@SenAmyKlobuchar & 5/20/19 & 3/3/20 & 6/20/20 \\
@WayneMessam & 5/20/19 & 12/2/19 & 9/25/20 \\
@sethmoulton & 5/20/19 & 11/14/19 & 9/25/20 \\
@BetoORourke & 5/20/19 & 11/14/19 & 9/25/20 \\
@TimRyan & 5/20/19 & 11/14/19 & 9/25/20 \\ 
@BernieSanders & 5/20/19 & - & -  \\ 
@ericswalwell & 5/20/19 & 11/14/19 & 9/25/20 \\
@ewarren & 5/20/19 & - & - \\ 
@SenWarren & 6/20/20 & - & - \\ 
@marwilliamson & 5/20/19 & - & -  \\ 
@AndrewYang & 5/20/19 & - & -  \\ 
@JoeSestak & 5/20/19 & 12/2/19 & 9/25/20 \\
@MikeGravel & 5/20/19 & 8/6/19 & 9/25/20 \\
@TomSteyer & 5/20/19 & - & - \\ 
@DevalPatrick & 5/20/19 & - & -  \\ 
@MikeBloomberg & 5/20/19 & - & -  \\ 
@staceyabrams & 6/20/20 & - & - \\
@SenDuckworth & 6/20/20 & - & - \\
@TammyforIL & 6/20/20 & - & - \\
@KeishaBottoms & 6/20/20 & - & -\\
@RepValDemings & 6/20/20 & - & - \\
@val\_demings & 6/20/20 & - & - \\
@AmbassadorRice & 6/20/20 & - & - \\
@GovMLG & 6/20/20 & - & - \\
@Michelle4NM & 6/20/20 & - & - \\
@SenatorBaldwin & 6/20/20 & - & - \\
@tammybaldwin & 6/20/20 & - & - \\
@KarenBassTweets & 6/20/20 & - & - \\ 
@RepKarenBass & 6/20/20 & - & - \\
@Maggie\_Hassan & 6/20/20 & - & - \\ 
@SenatorHassan & 6/20/20 & - & - \\
@GovRaimondo & 6/20/20 & - & - \\
@GinaRaimondo & 6/20/20 & - & - \\
@GovWhitmer & 6/20/20 & - & - \\
@gretchenwhitmer & 6/20/20 & - & - \\

    \end{tabular}
    \caption{A sample of the mentions and accounts that we actively tracked (v1.0 --- October 1, 2020).}
    \label{mention_table}
\end{table}

\begin{table}[t]
    \centering
    \footnotesize
    \begin{tabular}{cc}
    \textbf{Keywords} & \textbf{Tracked Since}\\
    \hline
ballot & 6/20/20 \\
mailin & 6/20/20 \\
mail-in &6/20/20\\
mail in & 6/20/20\\
donaldtrump & 9/12/20 \\
donaldjtrump & 9/12/20 \\
donald j trump &  9/12/20 \\
donald trump & 9/12/20 \\
don trump & 9/12/20 \\
joe biden & 9/12/20 \\
joebiden & 9/12/20 \\
biden & 9/12/20 \\
mike pence & 9/12/20 \\
michael pence & 9/12/20 \\
mikepence & 9/12/20 \\
michaelpence & 9/12/20 \\
kamala harris & 9/12/20 \\
kamala & 9/12/20 \\
kamalaharris & 9/12/20 \\
 trump & 9/13/20 \\
PresidentTrump & 9/13/20 \\
MAGA & 9/13/20 \\
trump2020 & 9/13/20 \\
Sleepy Joe & 9/13/20 \\ 
Sleepyjoe & 9/13/20 \\
HidenBiden & 9/13/20 \\
CreepyJoeBiden & 9/13/20 \\
NeverBiden & 9/13/20 \\
BidenUkraineScandal & 9/13/20 \\
DumpTrump & 9/13/20 \\
NeverTrump & 9/13/20 \\
VoteRed & 9/13/20 \\
VoteBlue & 9/13/20 \\
RussiaHoax & 9/13/20 \\
     \end{tabular}
    \caption{A sample of keywords that we actively tracked in our Twitter collection (v1.0 --- October 1, 2020).}
    \label{keyword_table}
\end{table}

\begin{table*}[t]
    \centering
    \begin{tabular}{cccc}
        \textbf{Conservative/Trump Campaign} & \textbf{Liberal/Biden Campaign} & \textbf{Conspiracy} & \textbf{Other}\\
        \hline
        MAGA & DemConvention & WWG1WGA & COVID19 \\
        Trump2020 & BidenHarris2020 & QAnon & coronavirus \\
        Trump & Biden2020 & Obamagate	& BlackLivesMatter \\
        RNC2020	& Democrats & & BLM \\
        KAG	& VoteBlueToSaveAmerica & & WalkAway \\
        MAGA2020 & JoeBiden & & BREAKING \\ 
        Trump2020Landslide & Biden & & Hydroxychloroquine \\
        AmericaFirst & WakeUpAmerica & & TrumpIsANationalDisgrace \\
        KAG2020	&  & & TrumpVirus\\		
        TulsaTrumpRally	&  & & TraitorTrump\\
        VoteRedToSaveAmerica &  & & TRE45ON \\		
         &  & & BountyGate\\		
	     &  & & TrumpIsALaughingStock\\		
        \hline

    \end{tabular}
    \caption{Top 35 hashtags (v1.0 --- October 1, 2020).}
    \label{tab:ht_table}
\end{table*}

\begin{table*}[t]
    \centering
    \begin{tabular}{cccc}
    
        \textbf{Conservative/Trump Campaign} & \textbf{Liberal/ Biden Campaign} & \textbf{Ballots} & \textbf{Other} \\
        \hline
        president @realdonaldtrump	& joe biden	& mail-in voting & law order\\
        donald trump & @joebiden @kamalaharris & mail-in ballots & law enforcement \\
        president trump & kamala harris & postal service & black lives\\
        fake news & & post office & white house \\
        @whitehouse @realdonaldtrump & & mail sorting & united states \\ 
        @realdonaldtrump @trump & & sorting machines & american people\\ 
        radical left & & & new york\\	
        @realdonaldtrump @foxnews& & & president united\\	
        @realdonaldtrump @potus & & & make sure\\	
        mr. president & & & god bless\\	
        @potus @realdonaldtrump	& & & executive order\\		
        @gop @realdonaldtrump & & & @realdonaldtrump @joebiden \\
        & & & vice president \\
        & & & four years \\
        & & & @hkrassenstein @realdonaldtrump \\ 
        & & & @itsjefftiedrich @realdonaldtrump \\
        & & & health care \\
        & & & many people \\
        \hline

    \end{tabular}
    \caption{The top 40 bigrams categorized by general topic (v1.0 --- October 1, 2020).}
    \label{tab:bigram_table}
\end{table*}
\section{Data \& Access Modalities}

\subsection{Release v1.0  (October 1, 2020)}
This initial dataset includes tweets collected from June 20, 2020 through September 6, 2020, containing \textbf{240,225,806} tweets in all. We note that this is only two months out of well over one year of data at our disposal as of the time of this writing. As we continue our computational efforts to pre-process and clean the rest of our existing dataset, we will be uploading batches of past and future data as they become available. The mentions/accounts and keywords that we follow can be found in Tables  \ref{mention_table} and \ref{keyword_table}, respectively. Furthermore, Tables \ref{tab:ht_table} and \ref{tab:bigram_table} show the top 35 most popular hashtags and bigrams in this dataset. Partisan trends emerge \cite{jiang2020political}, alongside with conspiracy theories \cite{ferrara2020types} and public heath related trends intertwined with COVID-19 \cite{chen2020tracking}.


\textbf{Access}:
The dataset is publicly available and continuously maintained on Github at this address: \textbf{{\url{https://github.com/echen102/us-pres-elections-2020}}}

The dataset is released in compliance with the Twitter's Terms \& Conditions and the Developer's Agreement and Policies.\footnote{\url{https://developer.twitter.com/en/developer-terms/agreement-and-policy}} 
This dataset is still presently being collected and will be periodically updated on our Github repository. Researchers who wish to use this dataset must agree to abide by the stipulations stated in the associated license and conform to Twitter's policies and regulations.

If you have technical questions about the data collection, please contact Emily Chen at \url{echen920@usc.edu}

\noindent If you have any further questions about this dataset please contact Dr. Emilio Ferrara at \url{emiliofe@usc.edu}


\bibliographystyle{aaai}
 \bibliography{references}
\end{document}